\documentclass[prl,twocolumn,superscriptaddress]{revtex4-1}

\usepackage[dvipdfmx]{graphicx}
\usepackage{dcolumn}% Align table columns on decimal point
\usepackage{bm}% bold math
\usepackage{color}

\begin{document}
%\preprint{APS/123-QED}
\title{180$^{\circ}$-phase shift of magnetoelastic waves observed by \\phase-resolved spin-wave tomography}

\author{Yusuke Hashimoto}
\affiliation{Advanced Institute for Materials Research, Tohoku University, Sendai 980-8577, Japan}

\author{Tom H. Johansen}
\affiliation{Department of Physics, University of Oslo, 0316 Oslo, Norway}
\affiliation{Institute for Superconducting and Electronic Materials, University of Wollongong, Northfields Avenue, Wollongong, NSW 2522, Australia}

\author{Eiji Saitoh}
\affiliation{Advanced Institute for Materials Research, Tohoku University, Sendai 980-8577, Japan}
\affiliation{Institute for Materials Research, Tohoku University, Sendai 980-8577, Japan}
\affiliation{Advanced Science Research Center, Japan Atomic Energy Agency, Tokai 319-1195, Japan}

\date{\today}

\begin{abstract}
We have investigated optically-excited magnetoelastic waves by phase-resolved spin-wave tomography (PSWaT).
PSWaT reconstructs dispersion relation of spin waves together with their phase information by using time-resolved magneto-optical imaging for spin-wave propagation followed by an analysis based on the convolution theorem and a complex Fourier transform.
In PSWaT spectra for a Bi-doped garnet film, we found a 180$^{\circ}$-phase shift of magnetoelastic waves at around the crossing of the dispersion relations of spin and elastic waves.
The result is explained by a coupling between spin waves and elastic waves through magnetoelastic interaction.
We also propose an efficient way for phase manipulation of magnetoelastic waves by rotating the orientation of magnetization less than 10$^{\circ}$.
\end{abstract}

\pacs{63.20.kk, 75.30.Ds, 75.40.Gb, 75.78.Jp}

%63.20.kk	Phonon interactions with other quasiparticles
%75.30.Ds	Spin waves (for spin-wave resonance, see 76.50.+g)
%75.40.Gb	Dynamic properties (dynamic susceptibility, spin waves, spin diffusion, dynamic scaling, etc.)
%75.78.Jp	Ultrafast magnetization dynamics and switching (for switching phenomena in ferroelectrics, see 77.80.Fm; for ultrafast spectroscopy, see 78.47.J-; for ultrafast processes in optics, see 42.65.Re)
%79.20.Ws	Multiphoton absorption (see also 82.50.Pt Multiphoton processes in photochemistry)

\maketitle

Spintronics is the research field aiming to develop novel devices based on spin degrees of freedom~\cite{Serga2010,Chumak2014a,Chumak2015}, which attracts great attention due to the potential for invention of new solid-state devices with low energy consumption~\cite{Chumak2014a} and a THz-working frequency~\cite{Bossini2017,Satoh:2017jg}.
In these devices, data may be transferred by spin waves, which are the collective excitation of the magnetization, ${\bf M}$.
So far, logic devices for NOT, XNOR, and NAND gates using the phase of spin waves have been demonstrated~\cite{Kostylev:2005fua,Schneider:2008fu}.

Recently, we developed spin-wave tomography (SWaT): reconstruction of dispersion relation of spin waves~\cite{Hashimoto:2017jb}.
SWaT is based on the convolution theorem, a complex Fourier transform (FT)~\cite{Gray:2012tq}, and a time-resolved magneto-optical imaging method for spin-wave propagation~\cite{Hashimoto2014}.
SWaT realizes the direct observation of dispersion relation of spin waves in a small-$k$ regime, so-called magnetostatic waves~\cite{Hashimoto:2017jb}.
We then developed an advanced application of SWaT, named phase-resolved SWaT (PSWaT)~\cite{Hashimoto:2018bz}.
PSWaT obtains the phase information of spin waves by separating the real and the imaginary components of the complex FT in the SWaT analysis.

Spin waves hybridized with elastic waves through magnetoelastic coupling are called magnetoelastic waves~\cite{Kittel:1958cv,Schlomann1960,Chikazumi1997,Dreher2012,Ruckriegel2014,Ogawa2015,Shen2015b,Hashimoto:2017tu,Shen:2018wf}.
The amplitude of magnetoelastic waves is resonantly enhanced at the crossing of the dispersion relations of spin waves and elastic waves, where the torque caused by elastic waves via magnetoelastic coupling is synchronized with the precessional motion of $\bf M$ in spin waves~\cite{Kittel:1958cv}.
The magnetoelastic waves have been observed by SWaT~\cite{Hashimoto:2017jb,Hashimoto:2017tu}, while then their phase information was disregarded.
In the phase of magnetoelastic waves, one can expect contributions by the phase of elastic waves, which are the source of magnetoelastic waves.

In this study, we investigated phase of magnetoelastic waves by PSWaT.
In the PSWaT spectra, we found 180$^{\circ}$ shift of the phase of magnetoelastic waves at around the crossings of dispersion relations of spin waves and elastic waves.
This feature is explained by a model based on the convolution theorem for optically-generated elastic waves and magnetoelastic coupling.
Finally, we propose an efficient way for the phase manipulation of magnetoelastic waves.

\begin{figure}
\includegraphics[width=8cm]{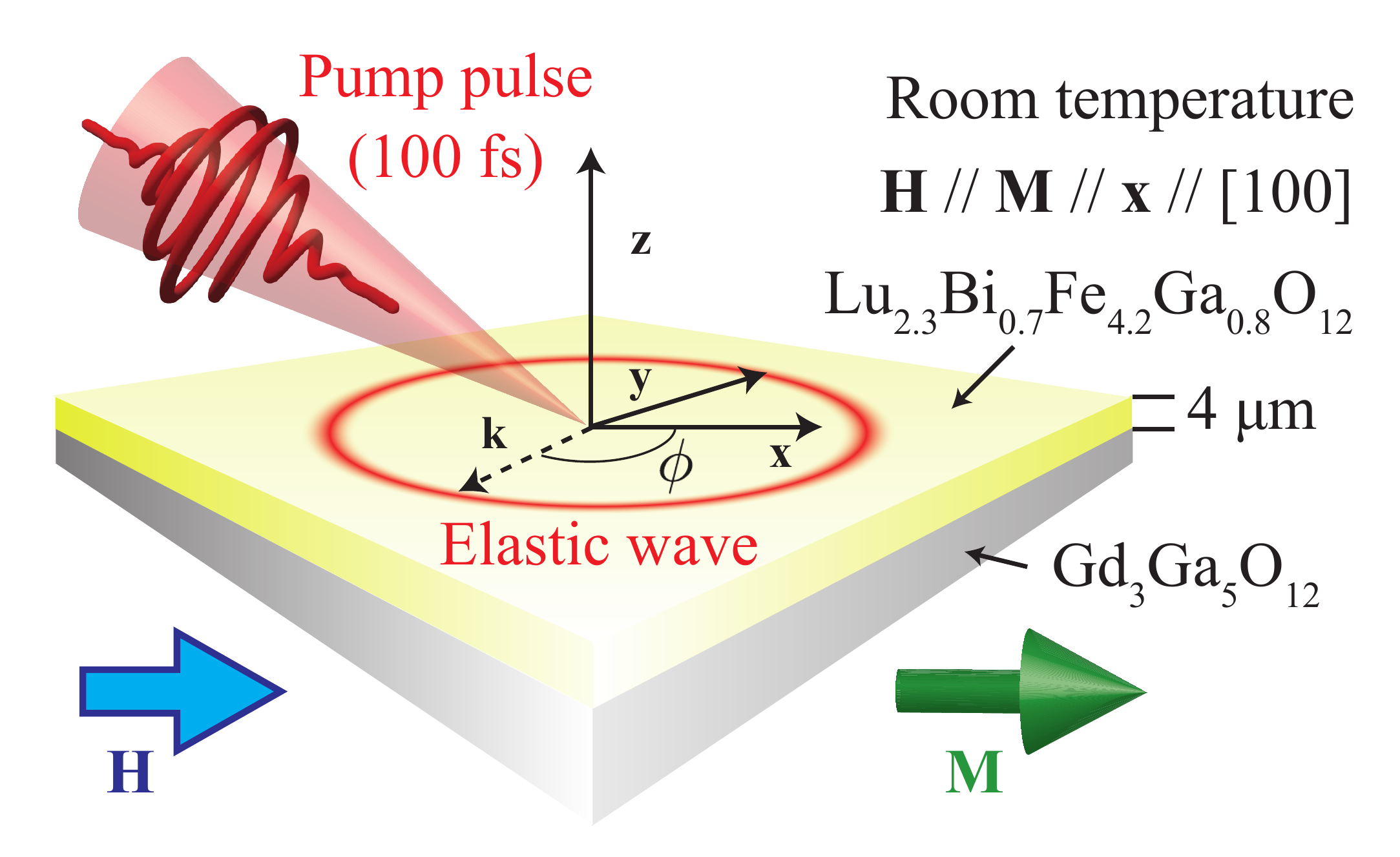}% Here is how to import EPS art
\caption{\label{FigExeSetup}
Schematic illustration of the experimental configuration and the rectangular coordinates ($x$, $y$, $z$).
The sample surface is in the $x$-$y$ plane with the x-axis along the orientation of ${\bf M}$, which is parallel to the [100] axis.
The orientation of ${\bf M}$ was controlled by an external magnetic field (${\bf H}$).
The sample normal is along the $z$-axis.
The pump pulse is focused on the sample surface at the origin of the coordinate to excite elastic waves, which is the source of the magnetoelastic waves investigated in this study.
The wavevector of elastic and magnetoelastic waves is denoted as $\bf k$.
The angle between $\bf k$ and $\bf M$ is defined as $\phi$.
}
\end{figure}

We used a 4-$\mu$m thick Lu$_{2.3}$Bi$_{0.7}$Fe$_{4.2}$Ga$_{0.8}$O$_{12}$ film grown on a Gd$_{3}$Ga$_{5}$O$_{12}$(001) substrate.
The film has a saturation magnetization of 780 Oe, which was aligned along the [100] axis by applying an external magnetic field of 240 Oe.

The propagation dynamics of optically-excited spin waves was observed with a time-resolved magneto-optical imaging system based on a pump-and-probe technique and a rotating analyzer method using a CCD camera~\cite{Hashimoto2014}.
A pulsed laser with the 800-nanometer center wavelength, 100-femtosecond time duration, and 1-kilohertz repetition frequency was used as a light source.
This beam was divided into pump and probe beams.
The center wavelength of the probe beam was changed to 630 nm, where the sample shows a large Faraday rotation angle (5.2$^{\circ}$) and a high transmissivity (41 $\%$)~\cite{Helseth2001,Hansteen2004}, by using an optical parametric amplifier.
The pump beam was circularly polarized and then tightly focused on the sample surface with a radius, $r_{0}$, of 1 $\mu$m.
The linearly-polarized probe beam was weakly focused on the sample surface with a radius of roughly 1 mm.
Optically-excited spin waves were observed with a time-resolved magneto-optical imaging system based on a rotating analyzer method using a CCD camera~\cite{Hashimoto2014}.
The spatial resolution of the obtained images is one micrometer, determined by the diffraction limit of the probe beam.
The time delay between the pump and the probe pulses was scanned from -1 ns to 13 ns.
All the experiments were performed at room temperature.
In Fig.~\ref{FigExeSetup}, we define the coordinates ($x, y, z$) and $\phi$, i.e. the angle between ${\bf M}$ and the wavevector, ${\bf k}$.
The details of the experimental setup were reported in Ref.~\onlinecite{Hashimoto2014}.

We investigated the excitation and the propagation dynamics of spin waves by SWaT~\cite{Hashimoto:2017jb, Hashimoto:2017tu} and PSWaT~\cite{Hashimoto:2018bz}.
Since spin waves were observed through the Faraday effect representing the magnetization along the sample depth direction [$m_{z}({\bf r}, t)$], the SWaT spectrum is denoted as $|m_{z}({\bf k},\omega)|$, where $\omega$ is an angular frequency.
The PSWaT spectrum is denoted as $m_{pqs}({\bf k},\omega)$, where $p$, $q$, and $s$ label the real ($r$) and imaginary ($i$) components of FT along the $x$, $y$, and time axes, respectively~\cite{Hashimoto:2018bz}.
The phase of spin waves determined by the temporal FT is defined by $\phi_{pq}({\bf k},\omega) = \tan^{-1}[m_{pqi}({\bf k},\omega)/m_{pqr}({\bf k},\omega)]$.

\begin{figure*}
\includegraphics[width=17cm]{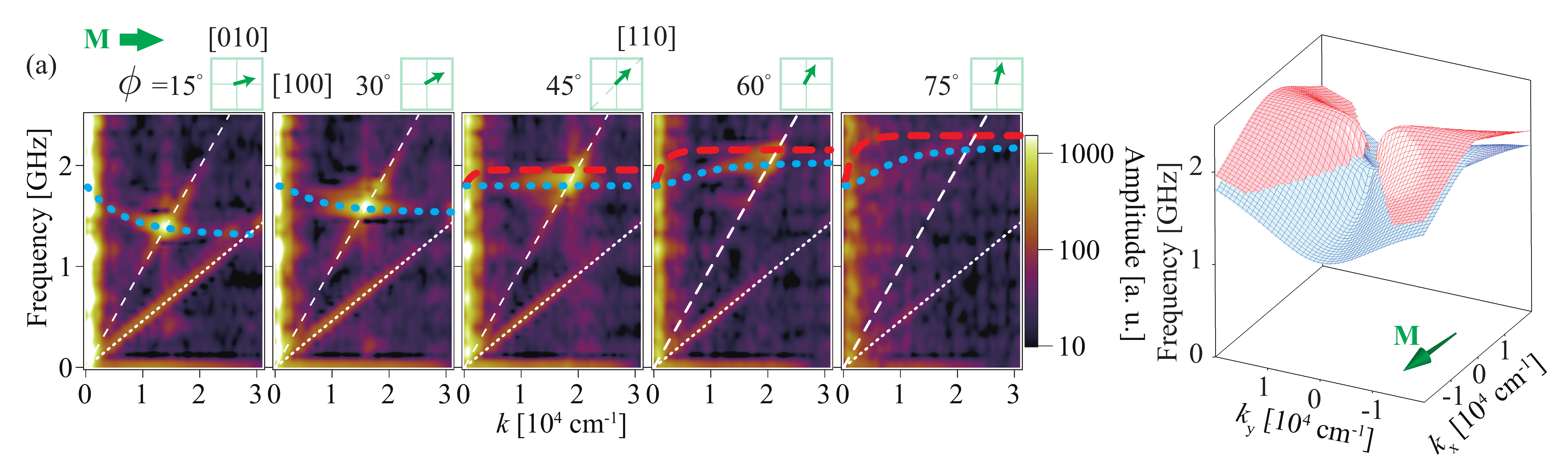}% Here is how to import EPS art
\caption{\label{FigSWaT}
(a) The angular dependence of the SWaT spectra for various ${\bf k}$ directions with an angle ($\phi$) between ${\bf M}$ and ${\bf k}$.
The color indicates the SWaT amplitude as shown in the color code.
The red dashed and the blue dotted lines are the dispersion relations of the surface and the volume modes of magnetostatic waves, respectively, calculated with the Damon-Eshbach theory~\cite{Hurben:1995fb} and the parameters obtained in Ref.~\cite{Hashimoto:2017jb}.
The white dashed and dotted lines are the dispersion relations of the longitudinal and the transversal modes of elastic waves, respectively.
(b) The dispersion relations of the surface (red plane) and the volume (blue plane) modes of the magnetostatic waves shown in (a).
The direction of $\bf M$ is indicated by the green arrow.
}
\end{figure*}

Let us first demonstrate the angular dependence of the SWaT spectra in Fig.~\ref{FigSWaT}(a).
The lines in Fig.~\ref{FigSWaT}(a) show the dispersion relations of spin waves and elastic waves calculated with the Damon-Eshbach theory~\cite{Hurben:1995fb} and the parameters obtained in Ref.~\cite{Hashimoto:2017jb}.
At around the crossing of the dispersion relations, we found signals ascribed to magnetoelastic waves.
By analyzing the obtained SWaT spectra, the dispersion relations of the volume and the surface modes of magnetostatic waves were reconstructed as shown in Fig.~\ref{FigSWaT}(b).

\begin{figure}
\includegraphics[width=8cm]{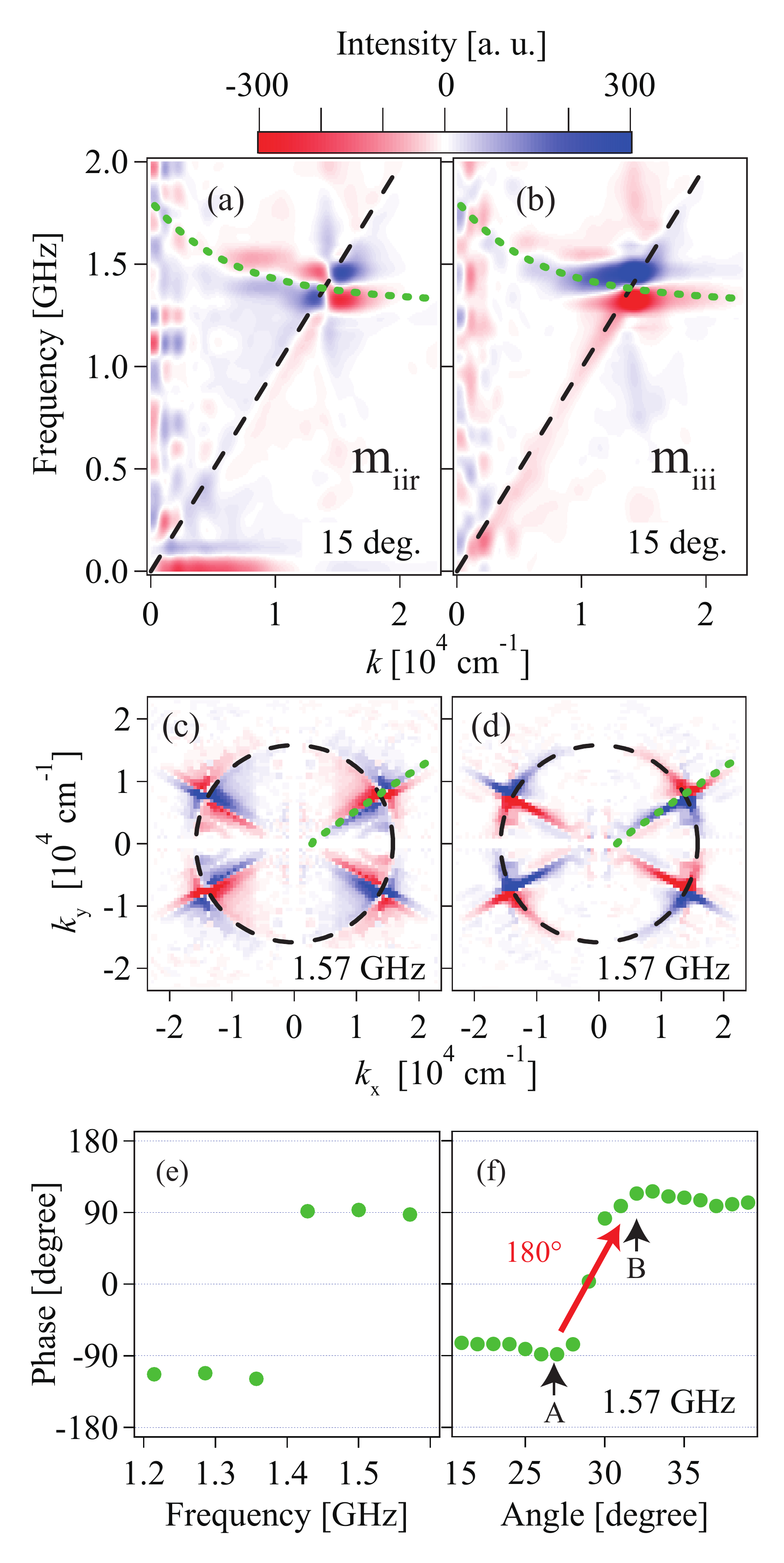}% Here is how to import EPS art
\caption{\label{FigPSWaTMEW}
(a, b) The cross-sectional views of the $m_{iir}$ (a) and the $m_{iii}$ (b) components of the PSWaT spectra along the $\phi$ = 15$^{\circ}$ direction.
(c, d) The cross-sectional views of the $m_{iir}$ (c) and the $m_{iii}$ (d) components of the PSWaT spectra at the frequency of 1.57 GHz.
In Figs.~(a)-(d), the color indicates the PSWaT intensity as shown in the color code.
The dispersion relation of the volume mode of magnetostatic waves calculated with the Damon-Eshbach model~\cite{Hurben:1995fb,Hashimoto:2017jb} is shown by the green dotted lines.
The dispersion relation of the longitudinal mode of elastic waves is shown by the black dashed lines.
(e) The frequency dependence of $\phi_{ii}$ along dispersion relation of elastic waves shown by the black dashed lines in (a) and (b).
(f) The angular dependence of $\phi_{ii}$ along dispersion relation of elastic waves shown by the black dashed lines in (c) and (d).
180$^\circ$-phase shift of magnetoelastic waves is obtained by rotating the orientation of the magnetization, which changes the phase of spin waves from the phase of point A to that of point B.
}
\end{figure}

Next, we demonstrate the observation of the phase of magnetoelastic waves by PSWaT.
In this study, we discuss the magnetoelastic waves generated by optically-excited longitudinal mode of elastic waves~\cite{Hashimoto:2017jb,Hashimoto:2017tu}.
This mode of magnetoelastic waves is observed in the $m_{iir}$ and $m_{iii}$ components of the PSWaT spectra representing signals having odd symmetries for both $x$ and $y$ axes [Figs.~\ref{FigPSWaTMEW}(a)-(d)]~\cite{Hashimoto:2018bz}.
Interestingly, we found that magnetoelastic waves show sign reversal at around the crossing of dispersion relations of spin waves and elastic waves.
At this point, the phase of magnetoelastic waves changes by 180$^{\circ}$ as shown in the plots of $\phi_{ii}$ in Figs.~\ref{FigPSWaTMEW}(e) and~\ref{FigPSWaTMEW}(f).
The explanation of the 180$^{\circ}$ shift of the phase of the magnetoelastic wave observed in the PSWaT spectra is the main topic of the following discussion.

Since the precession angle of ${\bf M}$ in the observed magnetoelastic waves is small (several degrees), the magnetic components of magnetoelastic waves may be written as ${\bf m} = \mbox{\boldmath $\chi$}{\bf h}$, where $\mbox{\boldmath $\chi$}$ is a dynamical susceptibility~\cite{Stancil:2009ux} and {\bf h} (= $h_x + ih_y$) represents an internal field 
applied to {\bf M} for the spin-wave excitation.
By solving the Landau-Lifshitz-Gilbert equation~\cite{Stancil:2009ux}, we write $\mbox{\boldmath $\chi$}({\bf k}, \omega) = \chi({\bf k}, \omega) + i\kappa({\bf k}, \omega)$ with $\chi({\bf k}, \omega) = \frac{\omega_{s}({\bf k}) \pm \omega}{(\omega_{s}({\bf k}) \pm \omega)^{2} + \alpha_{s}^{2}\omega^{2}}$ and $\kappa({\bf k}, \omega) = \frac{\alpha_{s} \omega}{(\omega_{s}({\bf k}) \pm \omega)^{2} + \alpha_{s}^{2}\omega^{2}}
$, where $\omega_{s}({\bf k})$ represents dispersion relation of spin waves, and $\alpha_{s}$ is the Gilbert damping constant.
%Here, we discuss magnetoelastic waves generated by the longitudinal mode of elastic waves through magnetoelastic coupling.
The torque caused by the longitudinal mode of elastic waves through magnetoelastic coupling can be treated as internal fields induced by elastic waves given by
$\mu_{0}h_{y}^{{\rm LM}}({\bf k},\omega) = b_{1}k\sin2\phi\mbox{\boldmath $\epsilon$}^{{\rm LM}}_{{\bf k}}({\bf k},\omega)$~\cite{Dreher2012}, where $\mbox{\boldmath $\epsilon$}^{{\rm LM}}_{{\bf k}}({\bf k},\omega)$ is the strain accompanied by the longitudinal mode of elastic waves along its ${\bf k}$ direction, and $b_{1}$ is the magnetoelastic coupling constant.
Because of the spatial symmetry of $h_{y}^{{\rm LM}}$, which has mirror symmetries for both $x$ and $y$ axes, the longitudinal mode of magnetoelastic waves appear in the $m_{iir}$ and $m_{iii}$ components of the PSWaT spectra.
This is consistent with our observations.

We have ascribed the optical excitation of such elastic waves to photo-induced charge transfer transition by two-photon absorption of the pump beam with the resonance at around 400 nm~\cite{Hashimoto:2017tu}.
All the waveforms observed in our experiments show strong magnetic field dependences and thus are attributed to spin waves.
The direct observation of elastic waves has not been obtained due to the limited sensitivity of our system.
In this study, we assume a response function of the optically-excited elastic waves to be 
\begin{eqnarray}
\label{eq:chielkt}
g({\bf r},t) = 2g_{0}\Theta(t)\sin\{\omega_{p}({\bf k})t - {\bf k}\cdot{\bf r}\}\exp(-\alpha_{p} t), 
\end{eqnarray}
where $g_{0}$ represents the strain accompanied by the longitudinal mode of elastic waves, $\Theta(t)$ is a Heaviside step function, $\omega_{p}({\bf k})$ is dispersion relation of elastic waves, and $\alpha_{p}$ represents the damping of elastic waves.
This assumption was employed to explain the phase of the magnetoelastic waves observed in the PSWaT spectra, although excitation mechanism of the elastic wave is out of the scope of this study.
With a model for the displacive excitation of coherent phonons~\cite{Zeiger:1992dea}, this assumption can be interpreted as the generation of elastic waves by a photoinduced change in the equilibrium position of lattice by photoexcited electrons~\cite{Wen:2013ka}.
With this assumption, the elastic waves excited by the illumination of the focused pump pulse via two-photon absorption may be given by
\begin{eqnarray}
\label{eq:chielkt}
\epsilon({\bf r},t) \propto \int\int\int g({\bf r} - {\bf a}, t - \tau)i_{p}({\bf a}, \tau)^2d{\bf a}d\tau
\end{eqnarray}
, where $i_{p}({\bf a}, \tau)$ represents the fluence of the pump pulse at the sample surface, assumed to be a Gaussian function with $i_{p}({\bf r},t) = \exp[-|r|^{2}/(2r_{0}^{2})]\delta(t)$.
A Dirac delta function in time [$\delta(t)$] was used since the duration of the pump pulse (sub-ps) is much shorter than the precession period of ${\bf M}$ ($\sim$ns).
Thus, the time-space FT of $g({\bf r}, t)$ gives ${\bf g}({\bf k},\omega) = g_{r}({\bf k},\omega) + ig_{i}({\bf k},\omega)$ with $g_{r}({\bf k},\omega) = g_{0}\frac{\omega \pm \omega_{p}({\bf k})}{[\omega \pm \omega_{p}({\bf k})]^2 + \alpha_{p}^2}$ and $g_{i}({\bf k},\omega) = g_{0}\frac{-\alpha_{p}}{[\omega \pm \omega_{p}({\bf k})]^2 + \alpha_{p}^2}$~\cite{Harris:2011ub}.
By using the convolution theorem~\cite{Harris:2011ub}, the time-space FT of $\epsilon({\bf r},t)$ is $\mbox{\boldmath $\epsilon$}({\bf k}, \omega) \propto i_{p}({\bf k},\omega)^2{\bf g}({\bf k}, \omega)$, where $i_{p}({\bf k},\omega) = \exp(-|k|^{2}r_{0}^{2}/2)$ is the time-space FT of $i_{p}({\bf r},t)$.

\begin{figure}
\includegraphics[width=8cm]{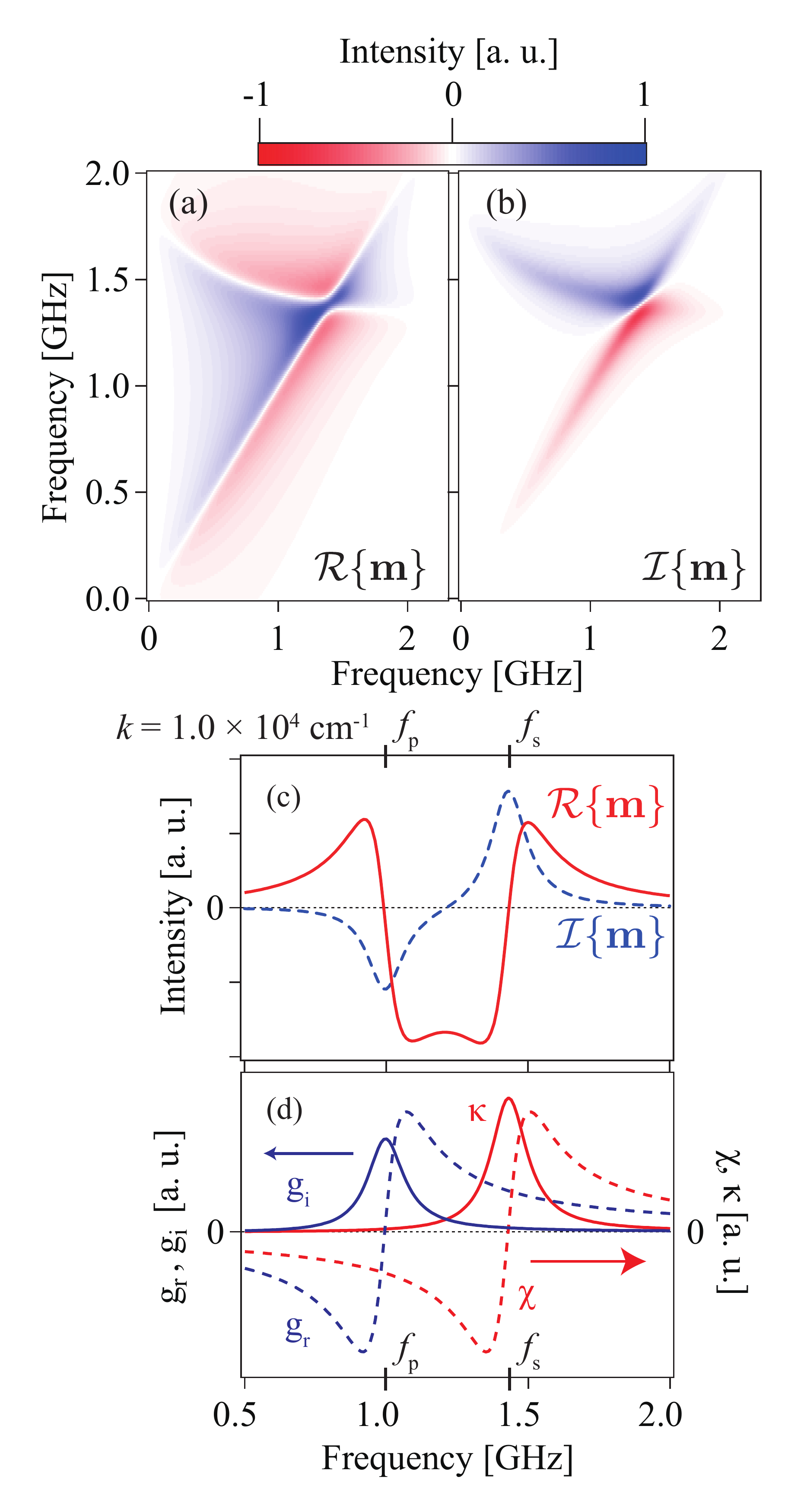}% Here is how to import EPS art
\caption{\label{FigMEWPhaseA}
(a) and (b) show the real ($\mathcal{R}$) and the imaginary ($\mathcal{I}$) components of {\bf m}, respectively, calculated with Eq.~\ref{eq:gpx1}.
We used $\omega_{s}({\bf k})$ and $\omega_{p}({\bf k})$ obtained by the SWaT spectra shown in Fig.~\ref{FigSWaT}(a) and $\alpha$ = 0.03, determined by the maximum time delay between pump and probe beams in our experimental setup ($T_{\rm{max}}$ = 13 ns), for both $\alpha_{s}$ and $\alpha_{p}$.
(c) The real and the imaginary components of {\bf m} at $k = 1.0 \times 10^{4}$ rad cm$^{-1}$ are shown by the red solid and the blue dashed lines, respectively.
(d) The spectra of $\chi$, $\kappa$, $g_{r}$, and $g_{i}$ used to calculate the data shown in (c). 
}
\end{figure}

We then calculate the PSWaT spectra of the longitudinal mode of magnetoelastic waves.
By using the equations shown above, we can write the longitudinal mode of magnetoelastic waves as 
\begin{eqnarray}
\label{eq:FTMEW}
{\bf m}^{{\rm LM}}({\bf k},\omega) \propto ib_{1}k\sin2\phi i_{p}({\bf k},\omega)^2{\bf g}^{{\rm LM}}({\bf k}, \omega)\mbox{\boldmath $\chi$}({\bf k},\omega).
\end{eqnarray}
Since $k$ and $i_{p}$ are real numbers while $\bf g$ and $\mbox{\boldmath $\chi$}$ are complex numbers, we write
\begin{eqnarray}
\label{eq:gpx1}
{\bf m} \propto ki_{p}^2{\bf g}\mbox{\boldmath $\chi$} & =& ki_{p}^2(g_{r} + ig_{i})(\chi +i\kappa) \nonumber \\
&=& ki_{p}^2[(\chi g_{r} - \kappa g_{i}) + i(\chi g_{i} + \kappa g_{r})].
\end{eqnarray}
The notation of $({\bf k},\omega)$ is omitted for clarity.
The real ($\mathcal{R}$) and the imaginary ($\mathcal{I}$) components of {\bf m} calculated with Eq.~\ref{eq:gpx1} are shown in Figs.~\ref{FigMEWPhaseA}(a) and~\ref{FigMEWPhaseA}(b), respectively.
Since the maximum time delay between the pump and probe pulses ($T_{\rm{max}}$ = 13 ns) is much shorter than the relaxation time of spin waves and elastic waves in garnet films, we used $\alpha$ limited by $T_{\rm{max}}$ giving $\alpha \sim $1/$(\omega T_{\rm{max}}) = 0.03$ for both $\alpha_p$ and $\alpha_s$.
In both $\mathcal{R}\{{\bf m}\}$ and $\mathcal{I}\{{\bf m}\}$, we can see the sign reversal of magnetoelastic waves at around dispersion relations of spin waves and elastic waves.
These trends are in good agreement with the experimental results shown in Figs.~\ref{FigPSWaTMEW}(a) and~\ref{FigPSWaTMEW}(b).
The sign reversal of the magnetoelastic waves are attributed to the sign reversal of $\chi$ and $g_{r}$ at the frequencies of the dispersion relations of spin waves ($f_{s}$) and elastic waves ($f_{p}$), respectively.
This is confirmed in the cross sections of $\mathcal{R}\{{\bf m}\}$ and $\mathcal{I}\{{\bf m}\}$ along at $k = 1.0 \times 10^{4}$ rad cm$^{-1}$ [Fig.~\ref{FigMEWPhaseA}(c)], calculated by Eq.~\ref{eq:gpx1} and $\chi$, $\kappa$, $g_{r}$, and $g_{i}$ [Fig.~\ref{FigMEWPhaseA}(d)].
By comparing Figs.~\ref{FigMEWPhaseA}(c) and~\ref{FigMEWPhaseA}(d), we find that the sign reversals in the data of $\mathcal{R}\{{\bf m}\}$ at $f_{s}$ and $f_{p}$ are caused by the sign reversal of $\chi$ and $g_{r}$ at $f_{s}$ and $f_{p}$, respectively.

Finally, we propose an efficient way for 180$^{\circ}$-phase manipulation of magnetoelastic waves.
Let us consider the case where magnetoelastic waves with ${\bf k}$ and $\omega$ at the point A in Fig.~\ref{FigPSWaTMEW}(f) are selectively excited by using, for instance, an interdigital transducer~\cite{Dreher2012}.
Then, by rotating $\bf M$ less than 10 $^{\circ}$, the phase of magnetoelastic waves is shifted from the phase of point A to that of point B in Fig.~\ref{FigPSWaTMEW}(f).
This realizes 180$^{\circ}$-phase manipulation of magnetoelastic waves.
Since the garnet film used in our experiments is magnetically soft to the magnetic field along the in-plane direction, we can easily rotate the orientation of the magnetization by rotating the direction of the external magnetic field.
Therefore, 180$^{\circ}$-phase manipulation of magnetoelastic waves by slightly rotating $\bf M$ may give us great potential for the development of future spin-wave devices.

In summary, we investigated phase of magnetoelastic waves in a Bi-doped garnet film by phase-resolved spin-wave tomography (PSWaT).
In the PSWaT spectra, we observed the 180$^{\circ}$-phase shift of magnetoelastic waves at around the crossing of the dispersion curves of spin and elastic waves.
This feature is consistent with a model based on the convolution theorem for spin waves excited by elastic waves through magnetoelastic coupling.

\begin{acknowledgments}
We thank Mr. T. Hioki, Dr. K. Sato and Dr. R. Ramos for fruitful discussions.
This work was financially supported by JST-ERATO Grant Number JPMJER1402, and World Premier International Research Center Initiative (WPI), all from MEXT, Japan.
\end{acknowledgments}

%\bibliography{Untitled}


\end{document}